# On the spatial resolution of EBSD in magnesium

Abhishek Tripathi[*], Stefan Zaefferer

[a]Max-Planck-Institute for Iron Research, Dusseldorf, D-40237, Germany

**Abstract**

We measured the *physical* lateral resolution of the electron backscatter diffraction (EBSD) technique for the case of pure magnesium and tungsten. Spatial resolution, among other parameters, depends significantly on the accelerating voltage and the atomic number of the material. For the case of lighter metals, it is supposed to be lower than in the case of heavier metals for a given accelerating voltage. In the present work, lateral resolution was measured in dependence of accelerating voltage on a straight high angle grain boundary which was positioned parallel (horizontal boundary) and perpendicular (vertical boundary) to the tilt axis of the specimen. For magnesium the best lateral resolution of 240 nm was obtained at an accelerating voltage of 5 kV. The resolution dramatically worsened to values as high as 3500 nm as the voltage was increased from 15 kV to 30 kV. The aspect ratio of horizontal and vertical lateral resolution tended to 1.0 at the accelerating voltage of 5 kV and to 2.5 at the accelerating voltage of 30 kV. These values as function of accelerating voltages were compared with those obtained on the high atomic number metal tungsten. Here resolution at 5 kV was about a quarter of that of magnesium. With increasing voltage, the value almost didn't change. For all voltages the resolution aspect ratio stayed close to 1.0.

## Introduction

Determination of crystallographic phase, crystal orientation, and further crystallographic measures by electron backscatter diffraction (EBSD) in scanning electron microscopy (SEM) have become a standard procedure and EBSD-based orientation microscopy has developed into one of the most versatile and powerful tools for microstructure evaluation [1,2]. Nevertheless, one important drawback of the technique remains, which is its spatial and lateral resolution which is only in the order of 100 nm for lateral resolution or $2 \cdot 10^4$ nm³ for the spatial volume resolution. For light materials (many minerals but also light metals like Al and Mg) the lateral resolution is even worse, obviously reaching into range of µm. For transmission electron microscopy (TEM), in contrast, the smallest volume that can be measured by spot or Kikuchi diffraction for orientation microscopy is about 10 times smaller and for transmission Kikuchi diffraction (TKD) performed on thin foils in the SEM the value is even still smaller. In order to improve the EBSD technique and make it even applicable for nanocrystalline and/or light materials, it is important to measure exact values and understand their dependence on measurement parameters.

In the case of EBSD, spatial resolution is significantly different for directions along the tilt axis and perpendicular to it, since the shape of the electron interaction volume is anisotropic. Spatial resolution furthermore rapidly worsens (that means its value becomes bigger) with increasing energy of the primary electron beam and with decreasing average atomic number of the target material. In general, the formation mechanisms of back scatter Kikuchi patterns consist of two steps. The first step is the incoherent scattering of the incoming primary beam electrons. Thermal diffuse scattering (TDS, aka phonon scattering) is known to be the primary mechanism involved in the first step [3,4]. The second step consists of elastic and coherent scattering (dynamic Bragg scattering) of electrons which leads to the formation of the Kikuchi pattern intensity distribution.

**Figure 1** shows an overview of the current state of literature available on the quantification of the spatial resolution in various metals ranging from Aluminum to Gold [3,5–8]. The values correspond to the physical spatial resolution [3] and not the effective lateral resolution [1,2].

Spatial resolution is measured based on the interaction volume overlap with a grain/phase/twin boundary, while effective resolution is determined by the extent to which the algorithms of indexing software can distinguish between patterns from adjoining grain/phase/twin boundaries. Generally, the value of effective spatial resolution is much smaller as compared to the physical spatial resolution. The distinction between physical and effective resolution is important as the effective resolution is only relevant in cases where well-distinguishable diffraction patterns, e.g. those obtained across a large angle grain boundary, are analyzed. In this case the effective resolution may lead to a significant and relevant improvement of the physical spatial resolution. In all other cases, for example when measuring orientation gradients inside of a grain, the physical spatial resolution is the relevant measure. Harland et al. [8] estimated the physical lateral resolution of 50 nm in gold (atomic number 79) at accelerating voltage of 30kV. Best effective lateral resolution for platinum with atomic number 78 was reported to be 6-9 nm at 25kV by Dingley [9]. For the case of medium atomic number ranged elements, Steinmetz and Zaefferer [10] measured physical lateral resolution of 30 nm and 10 nm at accelerating voltage of 15kV and 7.5kV respectively in twinning induced plasticity (TWIP) steel samples. For the case of aluminum, best resolution of 660 and 250nm for normal and parallel to the tilt axis respectively was reported in Baba-Kishi *et. al*. [5]. In the present work, we investigated the lateral resolution of EBSD for the case of light atomic weight element magnesium (atomic number 12, mass density 1.738 g cm$^{-3}$) as a function of varying accelerating voltages. Resolution was compared for the grain boundaries parallel and perpendicular to the tilt axis of the specimen. It was further compared with corresponding values for the case of a metal with a much higher atomic number, namely tungsten (atomic number 74, mass density 19.25 g cm$^{-3}$). For the present text, the term lateral resolution corresponds to the physical spatial resolution and not the effective resolution.

## Experimental methods

Pure cast magnesium samples were homogenized at the temperature of 500°C for the duration of 24 hours and immediately water quenched. Preparation of the homogenized samples were further done for EBSD measurements. Standard metallographic procedures were used which included grinding with 4000 grit SiC paper, polishing with diamond suspension of 1μm particle size, followed by electro-polishing in a 5:3 by volume ratio of solution of ethanol and ortho-phosphoric acid for 30 minutes. The potential difference of 2V was maintained during the polishing. The sample was rinsed with flowing tap water and immediately immersed in ethanol after the electro-polishing. The specimen was then polished with colloidal silica solution (0.05 μm particle size). The final step involved ion-milling in a Gatan$^{TM}$ PECS-682 system operated at 2 keV with 30 μA current at 10 RPM rotation speed at 75° tilt for the duration of 60 minutes. EBSD measurements were done on a Zeiss Crossbeam 1540 focused ion beam system with a 464 × 464 pixel Hikari camera. OIM data collection software by EDAX/TSL was used for pattern acquisition. A nearly straight and long grain boundary was located and high resolution EBSD measurements were performed across the boundary at varying accelerating voltages viz. 30, 15, 10 and 5 keV. A regular square grid was used in the present measurements. **Figure 2a** shows the grain structure in the inverse pole figure mapping after homogenization (500°C for 24 hours) depicted in the inverse pole figure mapping. The grain boundary selected for the orientation measurement is indicated with dotted lines in the figure with the grains labelled as A and B. This grain boundary is chosen since it's trace is parallel to the tilt axis of the specimen and also fairly straight. Next, the specimen was rotated in plane by 90°, so that the chosen boundary became perpendicular to the tilt axis of the specimen. As done earlier, high resolution EBSD measurements were performed again across the rotated boundary at same accelerating voltages. To avoid repeated measurements on electron beam contaminated areas, no two measurements were done on the exact same area, and hence the long grain boundary was initially selected. Working distance of 11 mm and step size of 50 nm was maintained for all the eight EBSD measurements described above. Magnification of 3500 was used at accelerating voltages of 10, 15 and 30 kV, at 5 kV magnification of 15000 was used. The pattern acquisition rate varied with the accelerating voltages. For the accelerating voltages of 30, 15, 10 and 5 kV pattern acquisition rates of 60, 20, 10 and 2 frames per second, respectively, were used. Camera settings and image processing parameters were kept same between all measurements. Kikuchi diffraction patterns were saved for each pixels at the different accelerating voltages and are shown in **Figure 3**. The Kikuchi patterns shown are the live patterns been captured at varying voltages and with the corresponding varying pattern acquisition speeds. The width of the Kikuchi bands increased with decreasing voltage.

## Results and Discussion

**Figures 2b** and **2c** show the grain boundary traces in the horizontal (parallel to the tilt axis of the specimen) and vertical position (perpendicular to the tilt axis of the specimen) and the corresponding electron beam interaction volume overlap geometry. The extent of overlap of the interaction volume of the electron beam is much larger in the horizontal case then in the vertical and hence the resolution is usually much poorer in the horizontal case than in the vertical case.

Kernel average misorientation (KAM) mapping [11] is used to check for the strains present in the grains as an estimate of the dislocation density near the grain boundary. Presence of a high density of dislocations near the grain boundaries can affect the resolution measurement, since it leads to further distortion in the electron backscatter Kikuchi diffraction patterns, which eventually worsen the resolution. The grains, however, show very low KAM values indicating presence of strain free regions throughout. The selected grain boundary is a high angle grain boundary with misorientation angle of 62.4°. Apart from dislocation density, inclination of the grain boundary plane with respect to the surface affects the Kikuchi pattern blurring near the grain boundary trace. The effect is more pronounced for steeply inclined boundaries. For observing the inclination of the grain boundary plane with the free surface, the grain boundary region was milled with $Ga^+$ ions in a focused ion beam instrument to expose the surface plane perpendicular to the free surface. However, because of the very low contrast in the back scatter electron (BSE) imaging mode, the inclination angle could not be determined accurately. A rough estimate of the inclination, however, can be used with image quality mapping across the grain boundary, which will be described subsequently. **Figure 4** shows a high resolution EBSD map of the grain boundary in horizontal position done at accelerating voltages of 30, 15, 10 and 5 kV. The maps are shown in notations of inverse pole figure mapping for the sample normal direction, confidence index mapping [12] and image quality mapping [13] to show the effect of the decreasing accelerating voltages on the detected thickness of grain boundaries. The confidence index is calculated by ranking of possible orientations (indexing solutions of the EBSD patterns) using a voting scheme for a given diffraction pattern. The formula for the confidence index is given as $I_c = (V_1 - V_2)/V_{total}$ where $V_1$ and $V_2$ are the number of votes for the first and second solutions and $V_{total}$ is the total possible number of votes. $I_c$ is 1 if only one solution is possible for a given pattern and it is 0 if the first and the second solutions are equally probable. The image quality value is an indication of the electron backscatter diffraction pattern quality. It is taken as the summation of the detected peaks in the Hough transform.

As can be seen in **figure 4**, at an accelerating voltage of 30kV, the grain boundary is poorly resolved. The detected thickness of the boundary decreases as the accelerating voltage is lowered, as is shown in both, the confidence index mapping and the image quality mapping. The sample was further in plane rotated by 90 degrees so that the specified grain boundary which was parallel to the tilt axis becomes perpendicular to the tilt axis of the sample and was further mapped at accelerating voltages of 30, 15, 10 and 5 keV.

**Figure 5** shows the vertical boundary in the inverse pole figure, confidence index and image quality mapping at the different accelerating voltages. As can be seen in the figure, with decreasing accelerating voltage the thickness of the detected grain boundary decreases, however, the thickness is much smaller as compared to the horizontal case, especially for the case of higher accelerating voltages. In addition, at lower accelerating voltage beam drift issue is observed indicated by the wavy appearance of the straight boundary. This issue is caused by the cycle of the objective lens cooling water temperature (about ±0.5°). The scanning speed is 30 times slower at the accelerating voltage of 5 keV compared to 30 keV. While it took about 2 h to measure the map at 5 kV and we counted 7 cooling cycles during this time, the same-sized map took only 4 minutes at 30 kV, containing not even one cooling cycle.

For quantitative estimation of the resolution, the image quality mapping is used. Any distortion of the crystal lattice in the interaction volume especially the diffracting volume affects the diffraction patterns and hence the image quality values. Grain boundaries as well are a special case of lattice distortion which can be mapped using image quality. As the electron beam in an EBSD map approaches a grain boundary, the back scattered electrons comes from a diffraction volume composed of crystal lattices from adjoining grains and hence a mix of diffraction patterns is observed. This leads to a decrease in the image quality values as the electron beam approaches the grain boundary. The image quality then increases again, as the beam goes farther from the grain boundary. This presence of a local minimum in the image quality value at the grain boundary can be used as an indicator of the resolution power. **Figure 6a** shows the schematic of the square grid EBSD mapping near the grain boundary trace using a step size of 50 nm. Image quality values across the grain boundary were extracted and an average of 15-line scan values was plotted with distance perpendicular to the boundary. However, in the representative schematic in **figure 6a** only 5 rows of the line scan are shown to convey the idea. The misorientation values along the line scans are mapped simultaneously and displayed by square symbols. Next, the values of image quality are indented to the point where the orientation changes. The orientation changes usually over one pixel where the voting algorithm detects the change of the pattern balance from one crystal to the other. Here this occurs over a 50 nm width, since the effective resolution value is much smaller compared with the physical spatial resolution. In contrast, the local minima in the image quality value spread over a significantly larger distance which is used here as the indication of the lateral resolution. Also, the degree of asymmetry in the values of image quality across the vertical grain boundary can give an indication of the inclination of the grain boundary plane. An asymmetric image quality values indicate a steeply inclined grain boundary while a more symmetric distribution indicates a boundary standing approximately perpendicular to the surface. For the present case, the values are not perfectly symmetric, but they are not too different as indicated in **figure 6b**: Tangents are drawn on the image quality values across the grain boundary at different sections as shown in **figure 6b**. The distance between the points of intersections of the tangents is determined and stated as the lateral resolution. In **figure 6b**, this methodology is shown for the vertical grain boundary mapped at the accelerating voltage of 15kV; the value of lateral resolution measured is 575 nm. This methodology is repeated for all the accelerating voltage for the vertical as well as horizontal boundaries. Alternatively, resolution may be defined as the distance between the points where the IQ signal has dropped to 80 % of the grain average IQ values. This leads to about 20 % smaller resolution numbers, here 460 nm.

**Figure 7a** shows the lateral resolution for the horizontal and vertical boundary as function of accelerating voltage for magnesium and tungsten. For the case of magnesium, the resolution ranges from 3500 nm to 240 nm at accelerating voltages of 30 kV to 5 kV respectively. The effect of the tilt of the boundary on the resolution becomes more pronounced with increasing accelerating voltage. The values for horizontal and vertical resolution approach similar values at lower accelerating voltage. This indicates that the shape of the interaction volume becomes less anisotropic as the accelerating voltage is lowered and vice versa. At 30kV, resolution dramatically worsens to as high as 3500 nm for the horizontal boundary and around 1400 nm for the vertical boundary. The ratio of horizontal and vertical resolution is near to 1.0 at lower voltages and increase to 2.5 at 30 kV. For the case of tungsten, the lateral resolution varies from 139 nm to 75 nm as the accelerating voltage varies from 30 kV to 5 kV for the case of horizontal boundary. The corresponding values for the vertical boundaries vary from 99 nm to 74 nm. **Table I** summaries the exact determined values for the different cases discussed here.

**Figure 7b** shows the ratio of the resolution with varying voltages. The resolution ratio for horizontal and vertical increases dramatically for the case of magnesium, however, the increase is much less for the case of tungsten. A schematic of the interaction volume of the primary electron beam with the sample in EBSD geometry is shown for accelerating voltage of 5 kV and 30 kV. The results show that the interaction volume becomes less anisotropic at the lower voltages. The interaction volume is much more anisotropic at higher accelerating voltage for the case of lighter metals. For the case of heavier metals, on the contrary, the interaction volume shape anisotropy is less and it also changes much less with accelerating voltages as compared to the much lighter metal namely Mg.

**Table I.** Measured values of horizontal and vertical resolution for Magnesium and Tungsten at different accelerating voltages.

| Accelerating Voltage (kV) | 30 | 15 | 10 | 5 |
|---|---|---|---|---|
| Horizontal Resolution (nm): Mg | 3593 | 738 | 345 | 250 |
| Vertical Resolution (nm): Mg | 1420 | 575 | 340 | 240 |
| Ratio: Mg | 2.53 | 1.28 | 1.01 | 1.04 |
| Horizontal Resolution (nm): W | 139 | 104 | 99 | 75 |
| Vertical Resolution (nm): W | 99 | 97 | 93 | 74 |
| Ratio: W | 1.40 | 1.12 | 1.02 | 1.01 |

The effect of the varying accelerating voltages on resolving power of a fine twin structure in magnesium is demonstrated in **Figure 8**. The fine tail of a tensile twin is mapped at different voltages. At 30 kV, the twins aren't visible as the resolution is too low. As the accelerating voltage is lowered to 15 kV, twins are resolved, however, the thickness of the twins can't be measured accurately. At 10 kV, the twins are properly resolved as well as the thickness can be measured with greater accuracy. At 5 kV, finally, twins can be resolved, however the conditions for EBSD are not ideal at lower accelerating voltages due to the low intensity of the detected back scattered electrons. This is in line with former findings of Humphreys at al. [1], who showed, at that time, that optimum resolution was obtained at 15 kV.

Measured resolution for the vertical and horizontal boundary as a function of accelerating voltage and atomic number is summarized in **Figure 9**. The measured values for the case of magnesium and tungsten in the present work is indicated by dotted lines. The resolution worsened much dramatically for the case of magnesium at higher accelerating voltage, compared with tungsten. As can be seen, much work has been done on resolution measurements for the transition metals, compared to metals with high and low atomic weight.

The voltage, U, and atomic number, Z, dependence of spatial resolution roughly follows the expected trend: spatial resolution becomes larger with increasing voltage and decreasing atomic number. This is due to the fact that thermal diffuse scattering, TDS (aka phonon scattering), is the main incoherent but quasi-elastic scattering mechanism involved in EBSD interaction volume formation. In earlier work [3] we reasoned, based on the theoretical treatment of Wang [4], that TDS is responsible for the angular distribution (max. between 10° and 120° with respect to the primary beam direction) and the energy distribution (a few eV to the primary beam energy, i.e. the zero-loss peak). Consequently we concluded that TDS also determines the spatial spatial resolution of the EBSD signal. Wang, in his textbook [14], presents an equation and reference data by Roussow [15], describing the mean free path length, $\Lambda$ of electrons scattered by TDS in dependence of Z and U. These data, however, show a quite different trend than ours: they roughly follow a logarithmic behaviour $\Lambda(U) \sim \ln(U)$, while ours show more a quadratic dependence $\Lambda(U) \sim U^2$ (note, both dependences are just estimated from the shape of the curves). This significant discrepancy may be, on one hand, due to the voltage range observed for both: while Roussow performs the calculations for U = 100 … 1000 kV we observe the behaviour for values between 5 and 30 kV. In this case the component of movement in the direction of the primary electron beam may become less significant relative to the components in other directions. This would also explain, that the interaction volume becomes more spherical when reducing the interaction volume. On the other hand, one has to take into account that not only the mean free path but also the mean scattering angle of the TDS events determines the spatial resolution. Wang [4,14] claims that the scattering angle is in the order of 2.5° for 200 kV and 10° for 15 kV, but the exact behaviour is not reported there. A proper estimation of the lateral resolution appears only possible by a Monte Carlo-type electron trajectory simulation using proper interaction cross sections for TDS. Unfortunately, all Monte-Carlo simulation programs available to us do not take TDS into account (which is visible, for example, by the fact that they do not simulate the zero-loss peak in the energy spectrum).

Overall, it can be clearly said that for increasing the resolution of the EBSD technique, lowering the accelerating voltage is a way out, however, it comes with the additional disadvantage of beam drift and poor indexing. This is particularly severe for magnesium which forms a relatively thick and not well conducting oxide film on its surface. Nevertheless, with improved new detectors, for example CMOS and direct electron detectors [16], this disadvantage can be overcome, which will be particularly beneficial for lower atomic number metals.

## Conclusions

In this study, a quantitative estimate of the lateral resolution of EBSD technique as a function of accelerating voltage for a light metal namely magnesium was done, using high resolution EBSD measurements on straight boundaries parallel and perpendicular to the tilt axis of a specimen in the EBSD set up. These measurements were complemented by similar measurements on a very heavy metal, namely tungsten. On magnesium best lateral resolution of 240 nm was measured at an accelerating voltage of 5 keV for the boundary perpendicular to the tilt axis. However, at lower voltages beam drift becomes more significant affecting the measurements. Lateral resolution worsens dramatically as the accelerating voltage is increased from 15 to 30 keV. The ratio of horizontal to vertical resolution values go from 2.53 at 30 keV to 1.04 at 5 keV, indicating a much lesser anisotropic interaction volume at the lower accelerating voltages. For tungsten, the minimum resolution is with 75 nm about a quarter of the value for magnesium, but it is not as small as we expected from literature values. Surprisingly, for tungsten resolution does not change much with acceleration voltage and also the shape of the interaction volume stays rather spherical. We interpret this by the fact that forward scattering becomes less important relative to lateral scattering for low voltages and high atomic numbers.

## Acknowledgements

The authors would like to acknowledge financial support from the German research foundation Deutsche Forschungsgemeinschaft (DFG) though priority programme SPP 1959, project No. 319282412.

## References

[1]   F.J. Humphreys, Grain and subgrain characterisation by electron backscatter diffraction, J. Mater. Sci. 36 (2001) 3833–3854.

[2]   F.J. Humphreys, Y. Huang, I. Brough, C. Harris, Electron backscatter diffraction of grain and subgrain structures - Resolution considerations, J. Microsc. 195 (1999) 212–216.

[3]   S. Zaefferer, On the formation mechanisms, spatial resolution and intensity of

backscatter Kikuchi patterns, Ultramicroscopy. 107 (2007) 254–266.

[4] Z.L. Wang, Thermal diffuse scattering in sub-angstrom quantitative electron microscopy - Phenomenon, effects and approaches, Micron. 34 (2003) 141–155.

[5] K.Z. Baba-Kishi, Measurement of crystal parameters on backscatter Kikuchi diffraction patterns, Scanning. 20 (1998) 117–127.

[6] T.C. Isabell, V.P. Dravid, Resolution and sensitivity of electron backscattered diffraction in a cold field emission gun SEM, Ultramicroscopy. 67 (1997) 59–68.

[7] S.X. Ren, E.A. Kenik, K.B. Alexander, A. Goyal, Exploring spatial resolution in electron back-scattered diffraction experiments via Monte Carlo simulation, Microsc. Microanal. 4 (1998) 15–22.

[8] C.J. Harland, C.J. Harland et. al., Proc. 9th Int. Cong. on 'Electron microscopy', Toronto, Ont., Canada, August 1978, Microscopical Society of Canada, 564 – 565, Proc. 9th Int. Cong. 'Electron Microsc. Toronto, Ont., Canada, Microsc. Soc. Canada. (1978) 564–565.

[9] D. Dingley, Progressive steps in the development of electron backscatter diffraction and orientation imaging microscopy, J. Microsc. 213 (2004) 214–224.

[10] D.R. Steinmetz, S. Zaefferer, Towards ultrahigh resolution EBSD by low accelerating voltage, Mater. Sci. Technol. 26 (2010) 640–645.

[11] M. Calcagnotto, D. Ponge, E. Demir, D. Raabe, Orientation gradients and geometrically necessary dislocations in ultrafine grained dual-phase steels studied by 2D and 3D EBSD, Mater. Sci. Eng. A. 527 (2010) 2738–2746.

[12] D.P. Field, Recent advances in the application of orientation imaging, Ultramicroscopy. 67 (1997) 1–9.

[13] S.I. Wright, M.M. Nowell, EBSD Image Quality Mapping, Microsc. Microanal. 12 (2006) 72–84.

[14] Z.L. Wang, Elastic and inelastic scattering in electron diffraction and imaging, 1995.

[15] C.J. Rossouw, P.R. Miller, J. Drennan, L.J. Allen, Quantitative absorption corrections for electron diffraction: Correlation between theory and experiment, Ultramicroscopy. 34 (1990) 149–163.

[16] K.P. Mingard, M. Stewart, M.G. Gee, S. Vespucci, C. Trager-Cowan, Practical application of direct electron detectors to EBSD mapping in 2D and 3D, Ultramicroscopy. 184 (2018) 242–251.

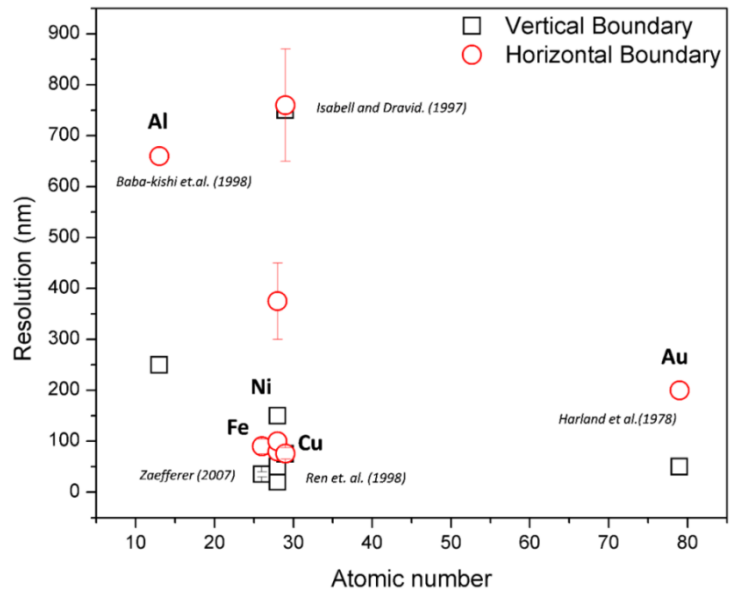

**Figure 1** Overview of the current literature on the resolution of metals for vertical and horizontal boundaries as function of their atomic number.

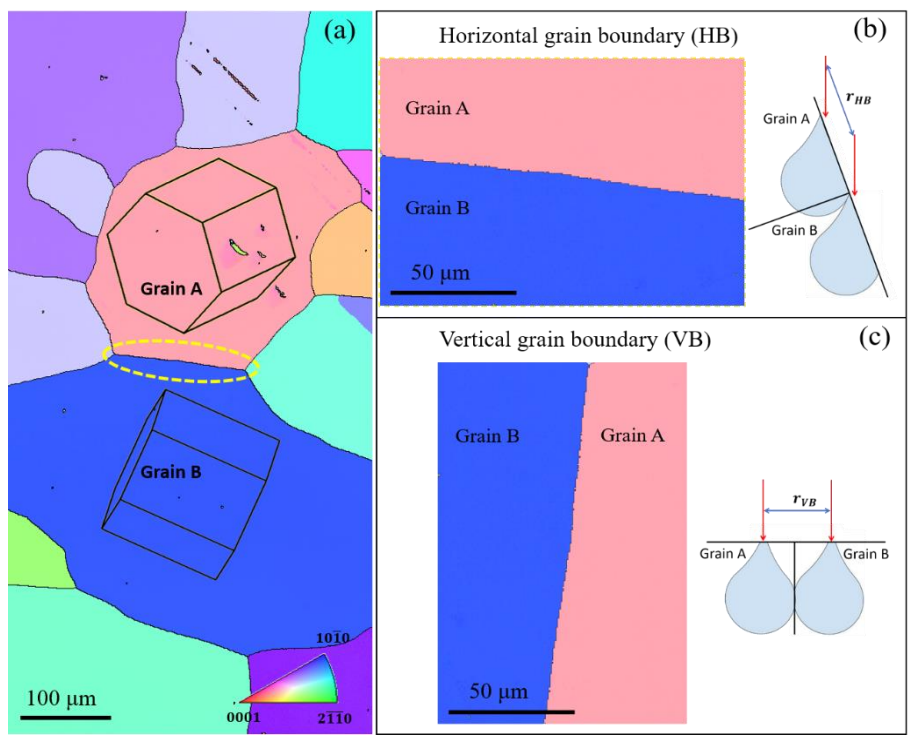

**Figure 2** a) Grain structure of as cast, pure and homogenized magnesium in the inverse pole figure notation. The boundary of interest is marked with the dotted lines with orientation of adjacent grains being depicted using the unit cells. Measurement is done at an accelerating voltage of 15kV.

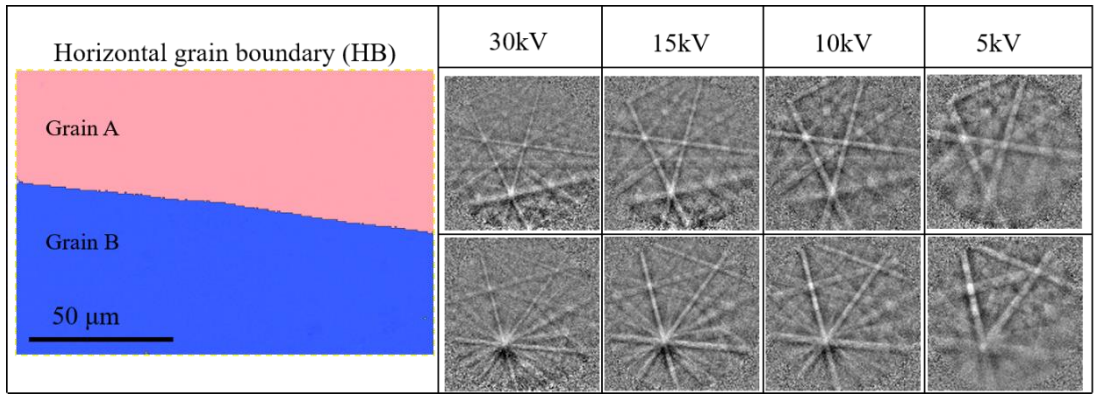

**Figure 3** Backscatter kikuchi patterns obtained at different voltages (viz. 30kV, 15kV, 10 kV and 5kV) at Grain A and Grain B. The patterns in first and second row are from Grain A and Grain B respectively and also are from regions away from the grain boundary.

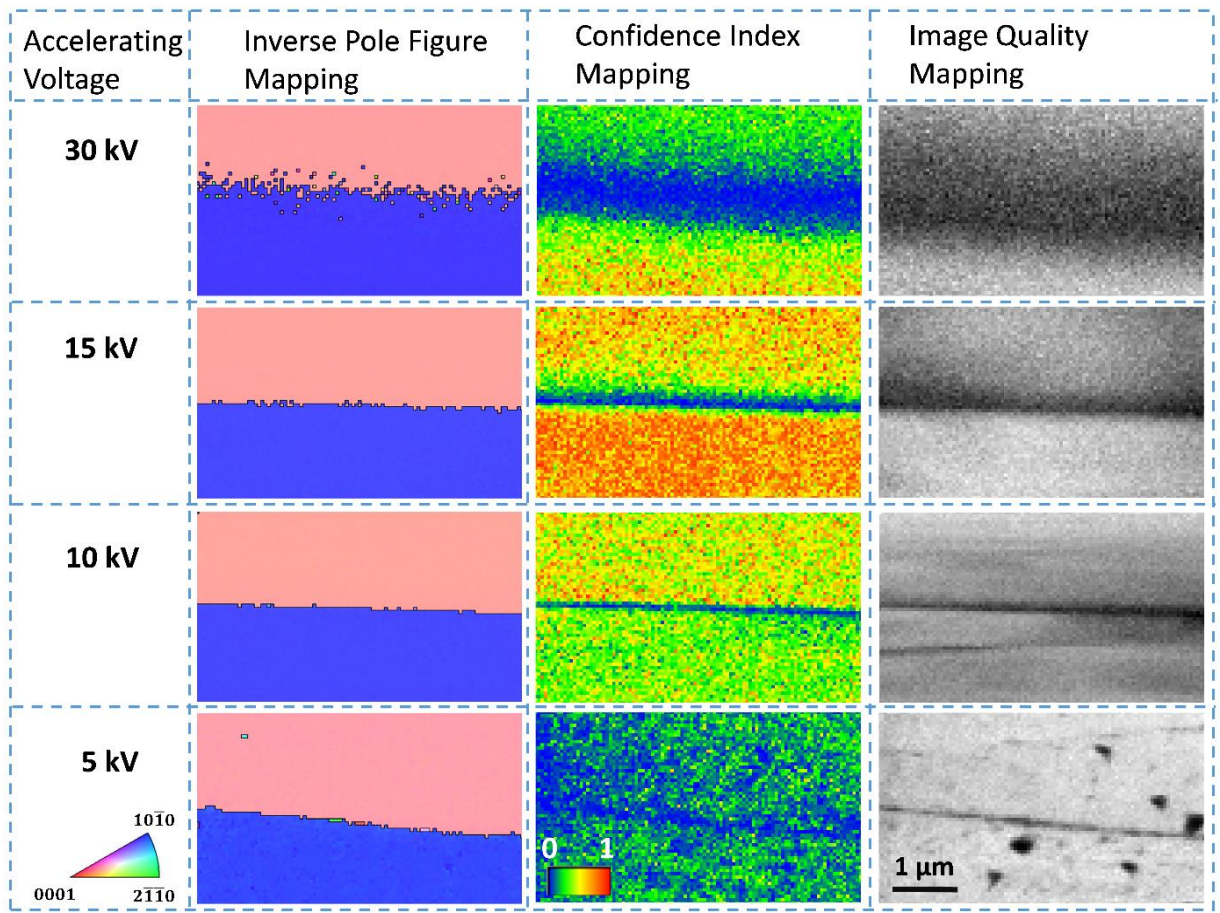

**Figure 4.** Inverse pole figure mapping, Confidence index mapping and Image quality mapping of the horizontal grain boundary at various accelerating voltages namely 30 kV, 15 kV, 10 kV and 5 kV.

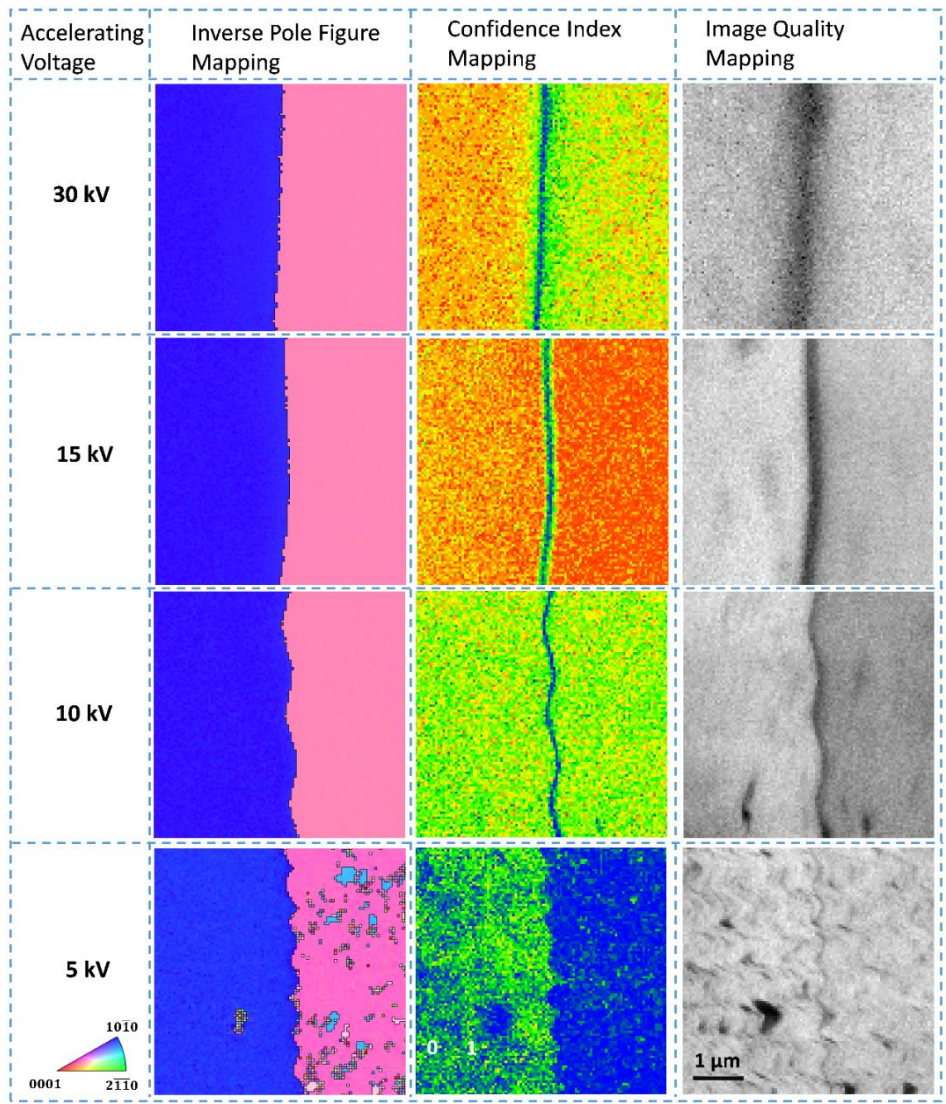

**Figure 5**. Inverse pole figure mapping, Confidence index mapping and Image quality mapping of the vertical grain boundary at various accelerating voltages namely 30 kV, 15 kV, 10 kV and 5 kV.

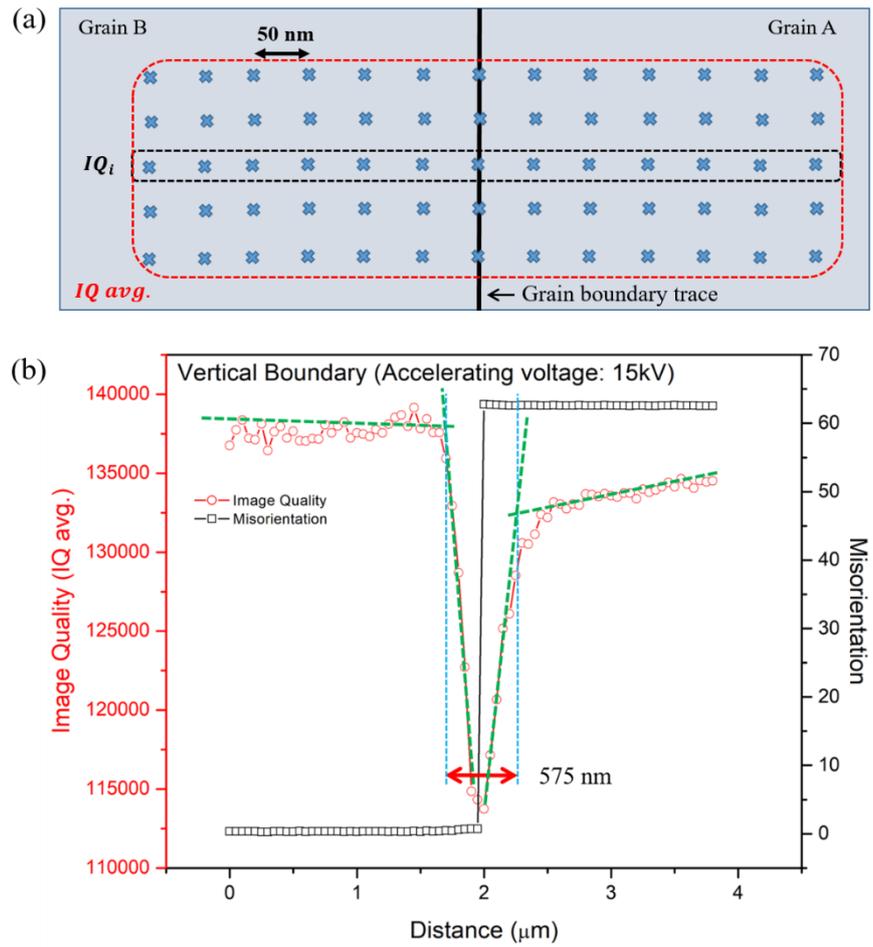

**Figure 6.** Example of the methodology used to estimate the resolution with the image quality values across the grain boundary. The example shown is the image quality values for the vertical grain boundary measured at the accelerating voltage of 15 kV.

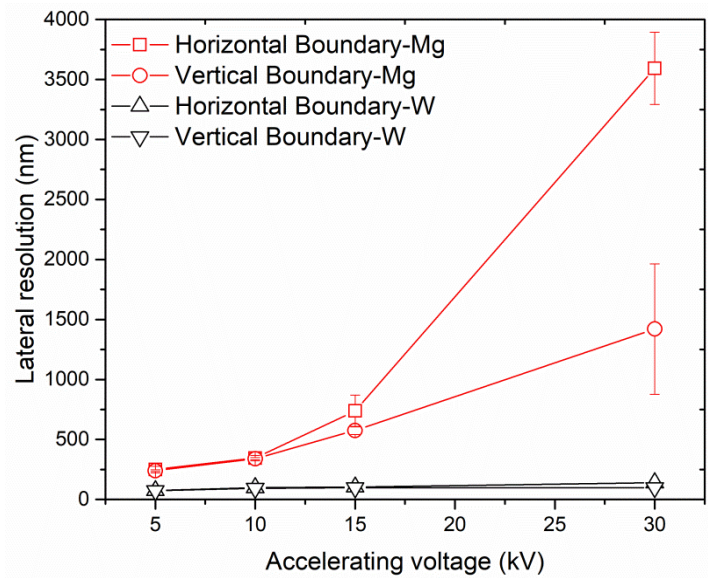

(a)

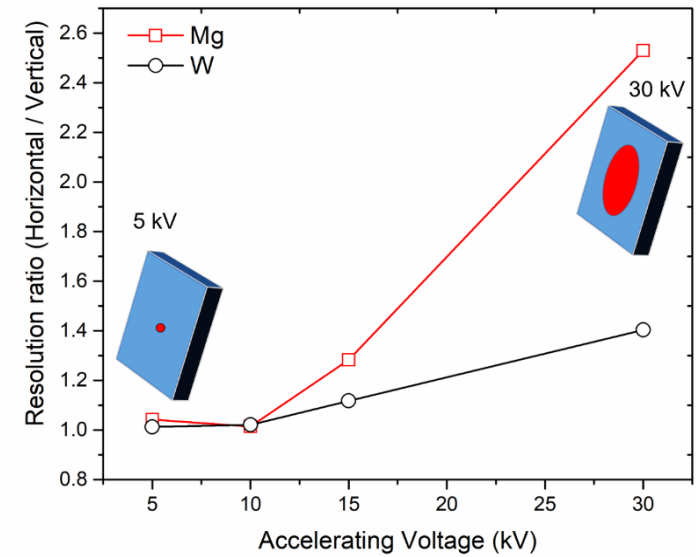

(b)

**Figure 7**. (a) Measured resolution for magnesium and tungsten for the vertical and horizontal grain boundary at the various accelerating voltages. (b) Ratio of the horizontal and vertical boundary resolutions for magnesium and tungsten. Also shown is the schematic of the top view of the interaction volume at accelerating voltage of 5kV and 30kV for the case of magnesium.

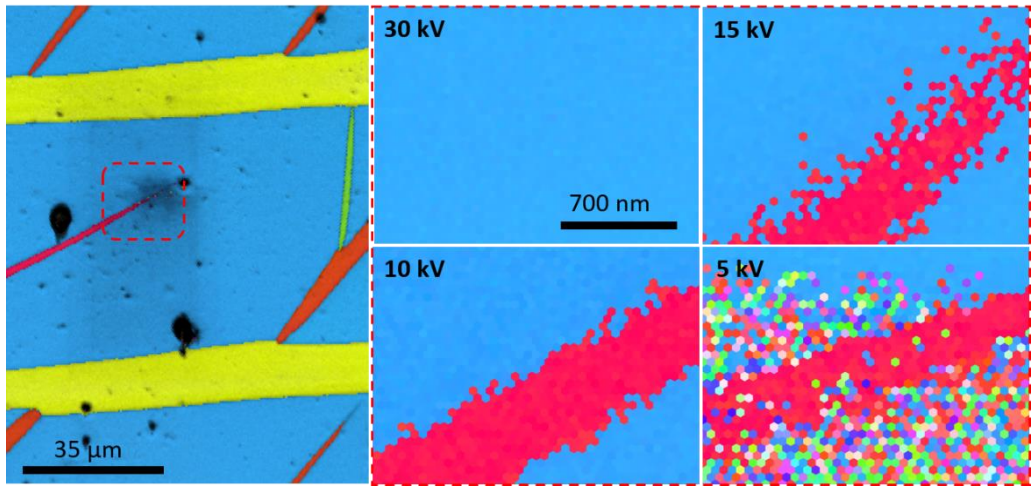

**Figure 8**. Comparison of the EBSD measurements of twins in pure magnesium at accelerating voltages of 30, 15, 10 and 5 kV.

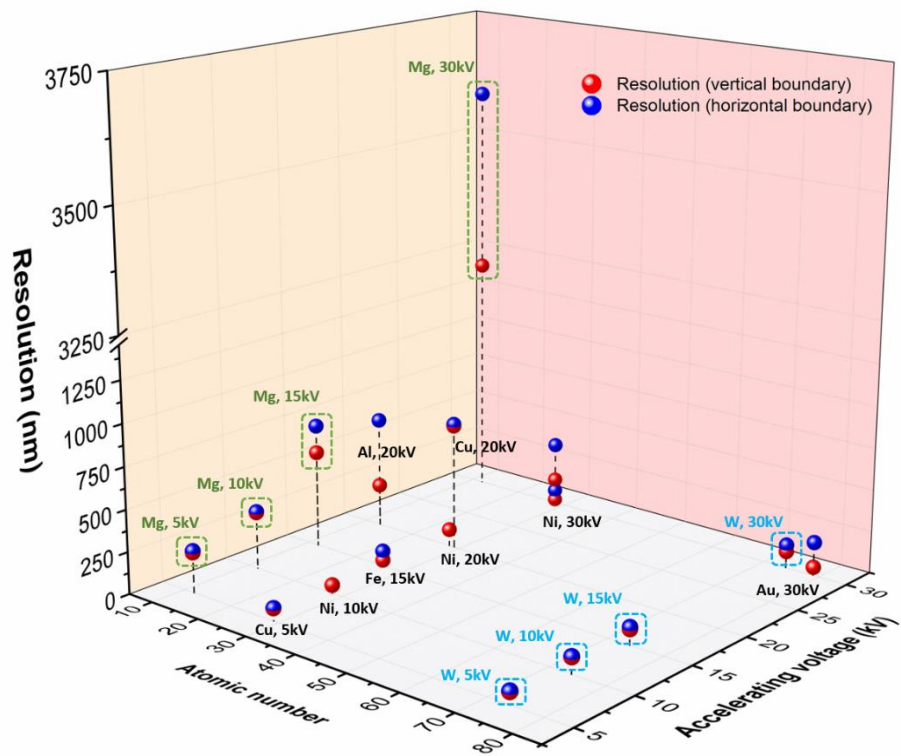

**Figure 9**. Overview of the current literature on the resolution of metals for vertical and horizontal boundaries with different accelerating voltages as function of their atomic number. This includes the resolution in case of magnesium and tungsten from the current work at accelerating voltages of 30, 15, 10 and 5 kV.